# Nonlinear Self-Calibrated Spectrometer with Single GeSe-InSe Heterojunction Device

Rana Darweesh,[1,2,†] Rajesh Kumar Yadav,[1,2,†] Elior Adler,[1,2] Michal Poplinger,[1,2] Adi Levi,[1,2] Jea-Jung Lee,[3] Amir Leshem,[1] Ashwin Ramasubramaniam,[4,5] Fengnian Xia,[3] and Doron Naveh[1,2,*]

[1] Faculty of Engineering, Bar-Ilan University, Ramat-Gan, Israel 52900.

[2] Institue of Nanotechnology and Advanced Materials, Bar-Ilan University, Ramat-Gan, Israel 52900.

[3] Department of Electrical Engineering, Yale University, New Haven, CT, USA.

[4] Department of Mechanical and Industrial Engineering, University of Massachusetts, Amherst, MA 01003, USA.

[5] Materials Science and Engineering Graduate Program, University of Massachusetts, Amherst, MA 01003, USA.

[†] These authors contributed equally. [*] Corresponding author: doron.naveh@biu.ac.il

**Optical spectroscopy – the measurement of electromagnetic spectra – is fundamental to various scientific domains and serves as the building block of numerous technologies.[1–3] Computational spectrometry is an emerging field that employs an array of photodetectors with different spectral responses or a single photodetector device with tunable spectral response, in conjunction with numerical algorithms, for spectroscopic measurements.[4–6] Compact single photodetectors made from layered materials are particularly attractive,[7–9] since they eliminate the need for bulky mechanical and optical components used in traditional spectrometers and can easily be engineered as heterostructures to optimize device performance.[7,10] However, compact tunable photodetectors are typically nonlinear devices and this adds complexity to extracting optical spectra from the device response. Here, we**

**report on the training of an artificial neural network (ANN) to recover the full nonlinear spectral photoresponse - a nonlinear problem of high dimensionality - of a single GeSe-InSe p-n heterojunction device. We demonstrate the functionality of a calibrated spectrometer in the spectral range of 400-1100 nm, with a small device footprint of $\sim 25 \times 25\ \mu m$, and we achieve a mean reconstruction error of $2 \times 10^{-4}$ for the power-spectrum at a spectral resolution of 0.35 nm. Using our device, we demonstrate a solution to metamerism, an apparent matching of colors with different power spectral distributions, which is a fundamental problem in optical imaging.**

Optical sensing, the measurement of the properties of light such as its spectrum, polarization, and power, is central to several fields of science and technology.[11,12] Accurate optical spectrometers with analytical calibration and resolution are typically expensive table-top machines containing moving optical components.[11,12] In contrast, computational optical spectroscopy[7–9] and sensing,[13,14] which entails the use of an algorithm and either a single on-chip tunable detector or an array of detectors to replace the optical components in conventional sensing instruments, allows for spectra[9] and polarization[15] measurements efficiently and compact manner. The operational principle of a computational spectrometer based on a single tunable device exploits the electrical tunability of the photodetector, *e.g.*, via voltage bias, for collecting the output photocurrent within a high-dimensional vector space defined by the number of variable input conditions. This process is termed an encoding process and, so far, it has been applied only to linear detectors since the non-linear problem was considered intractable. Linear computational detectors exploit the relation $I_{ph}(V) = \int_{\lambda_1}^{\lambda_2} R(V,\lambda)\mathcal{P}(\lambda)\, d\lambda$, where $R$ is the responsivity, $\mathcal{P}(\lambda)$ is the spectral power density to be measured, $I_{ph}(V)$ is the photocurrent, and $\lambda$ is the wavelength.[9,16] The voltage-dependent responsivity matrix, $R(V; \lambda)$,[7,9,10] operates as a transformation operator that maps the measured photocurrent $I_{ph}(V)$ on to the spectrum $\mathcal{P}(\lambda)$. These linear relations between the responsivity, current and spectrum allow for training a transformation matrix that connects the measured photocurrent to the unknown spectrum, via linear regression techniques. [7,9,10] However, most semiconductor devices can also operate in the nonlinear regime. [17,18] For example, diodes, field-effect transistors, and bipolar transistors often display a nonlinear response that is considered disadvantageous for high-fidelity analog and digital communication systems[19,20] and in optical spectrometers.[21] In the case of nonlinear photoresponse, the existence of a mapping between the photocurrent and spectrum is not guaranteed, and if such a mapping exists, it must rely on a larger parameter space to account for the nonlinearity. In this work, we use a voltage-tunable heterojunction device of p-GeSe/n-InSe (see **Figure 1a**) that is amenable to tuning of both the spectral response and its higher-order nonlinearities. To resolve the complex mapping transformation between the spectrum and photocurrent, we encode the nonlinear response of the device by training an ANN (**Figure 1b-1c**) that captures device's response as a function of bias voltage and spectrum. Unknown spectra can then be analyzed by decoding the voltage-dependent photocurrent response of the device (**Figure 1d**) and performing the inverse mapping via the trained ANN (**Figure 1e**) to reconstruct the power spectrum (**Figure 1f**).

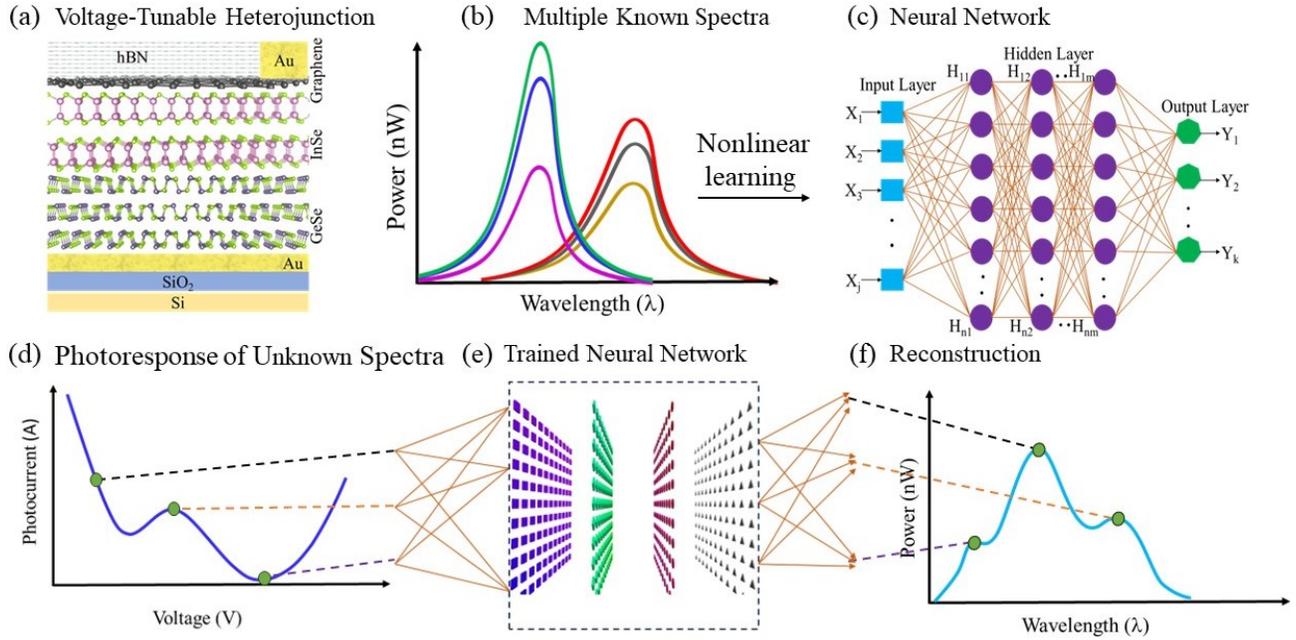

**Figure 1.** Illustration of the nonlinear learning and reconstruction process: Schematic atomistic representation of vertical InSe/GeSe device (a), that is exposed to known power-modulated spectra (b), for training an artificial neural network that captures the device's nonlinear response (c). A measured photocurrent vector of an unknown spectrum (d) is then analyzed with the trained neural network (e), enabling the reconstruction of the unknown spectrum (f).

The voltage-tunable InSe/GeSe heterojunction device (**Figure 2a**) consist of a stack of ~4 layers of p-type GeSe and ~7 layers of n-type InSe; the GeSe side is in contact with a gold electrode while the InSe side is in contact with transparent graphene electrode, and the entire device is encapsulated in hBN (**Figure 1a** and **Supplementary Figure 1**). The Raman spectrum corresponding to this structure is presented in **Figure 2b** (and **Supplementary Figures 2 and 3**), and the current-voltage transfer curves confirm the formation of a p-n junction (**Figure 2c**; also see SI for results of a lateral device as a control experiment). While the principle of voltage-tunable band alignment and spectral response in 2D heterostructures is known,[7,10] the nonlinear response of the GeSe-InSe device is not yet understood. Previous reports on the nonlinear optical response of these materials have dealt primarily with second-harmonic generation, originating from their non-centrosymmetric crystal structure. [22–24] In the present case, the tunable nonlinear response of the heterostructure device arises from interfacial charge transfer at the p-n junction. When bias voltage is applied to this heterojunction, the built-in potential is modified as well as the optical polarizability associated with the observed voltage and nonlinear spectral response (**Figure 3**). A general model describing the nonlinear response of the device may be written as

$$I_V = \sum_\lambda R_{V,\lambda} \mathcal{P}_\lambda^{\gamma(V,\lambda)}, \begin{cases} \lambda \in [\lambda_1 \dots \lambda_N] \\ V \in [V_1 \dots V_M] \end{cases}, \quad (1)$$

where the nonlinear coefficients $\gamma(V,\lambda)$ are unknown and span a high-dimensional $N \times M$ parameter space, $I_V$ is the photocurrent at voltage $V$, $R_{V,\lambda}$ is the responsivity at voltage $V$ and wavelength $\lambda$, and $\mathcal{P}_\lambda$ is the component of the power spectrum at wavelength $\lambda$. The dependence of our device's nonlinear response on the applied voltage and spectrum is evaluated from the measurements in **Figure 3**.

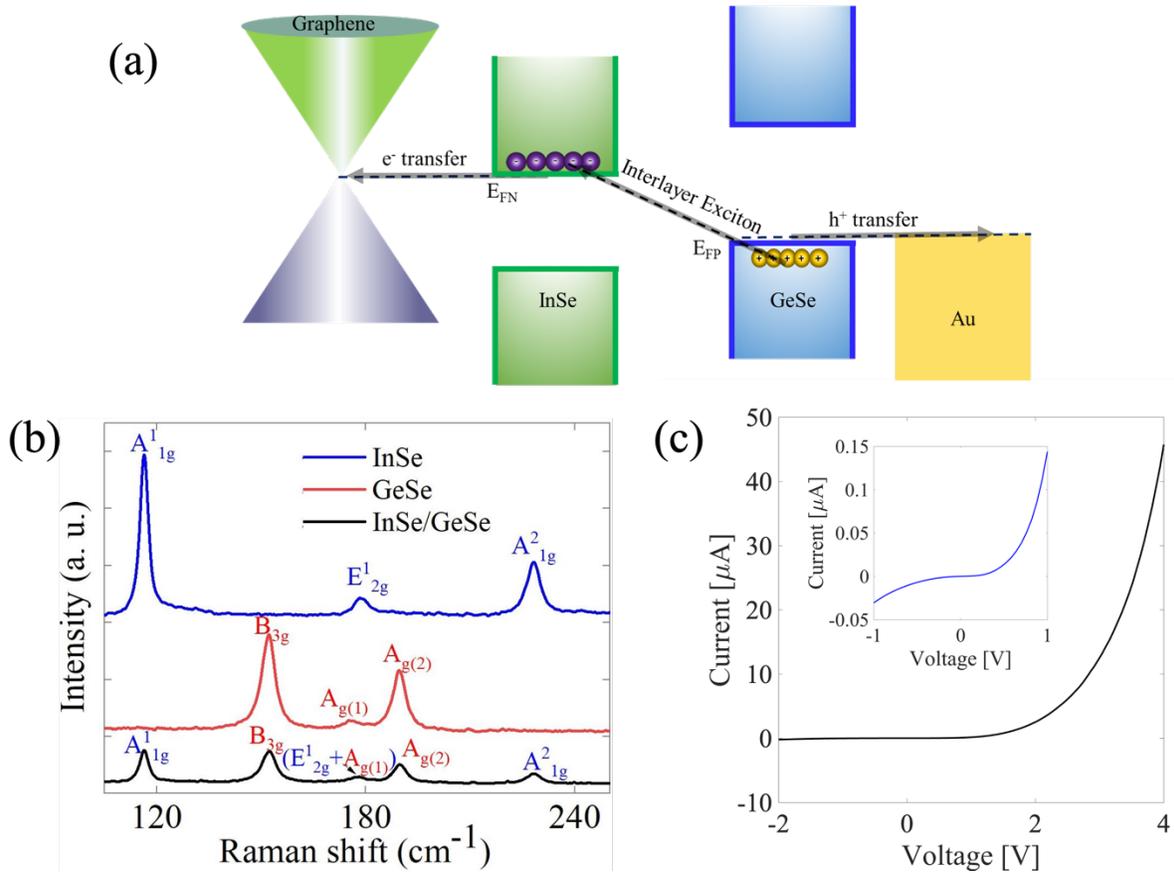

**Figure 2.** A schematic depicting the voltage-adjustable band alignment of InSe/GeSe (a) and the corresponding Raman spectra of InSe, GeSe and InSe/GeSe, respectively (b). The device current-voltage transfer curve, with the inset displaying voltage range of the photoresponse measurements (c).

The InSe/GeSe heterojunction device was illuminated by seven different light sources, with each source being power-modulated at ten different intensities (**Figures 3a, 3b**) while covering the spectral range of 400-1100 nm, and the resulting voltage-dependent photocurrents recorded, as shown in **Figures 3c** and **3d**. From these power-dependent $I_{ph}(V)$ data, the photocurrent-power relation is retrieved for each fixed voltage (**Figures 3e, 3f**) and is fit to a saturating nonlinear model, $I_{ph} = RP^\gamma$, where the total power is defined by $P = \int_0^\infty \mathcal{P}(\lambda)d\lambda$. Eventually, the voltage-dependent nonlinear coefficients, $\gamma(V, \lambda)$, are evaluated for each spectrum (**Figure 3a, 3b**) and, as seen from **Figure 3g** and **Figure 3h**, depend strongly on both the spectrum and voltage. In this work, we measured the device response for $N = 4,000$ wavelengths and $M = 101$ voltages, after which we interpolated the data to $N = M = 2,000$ points (see methods for details). This leads to a set of $D_{NM} = 4 \times 10^6$ nonlinear coefficients, according to the model in Equation (1). This large number of free independent parameters adds significant computational complexity as compared to previously studied linear response models. [6,7,10,25–27]

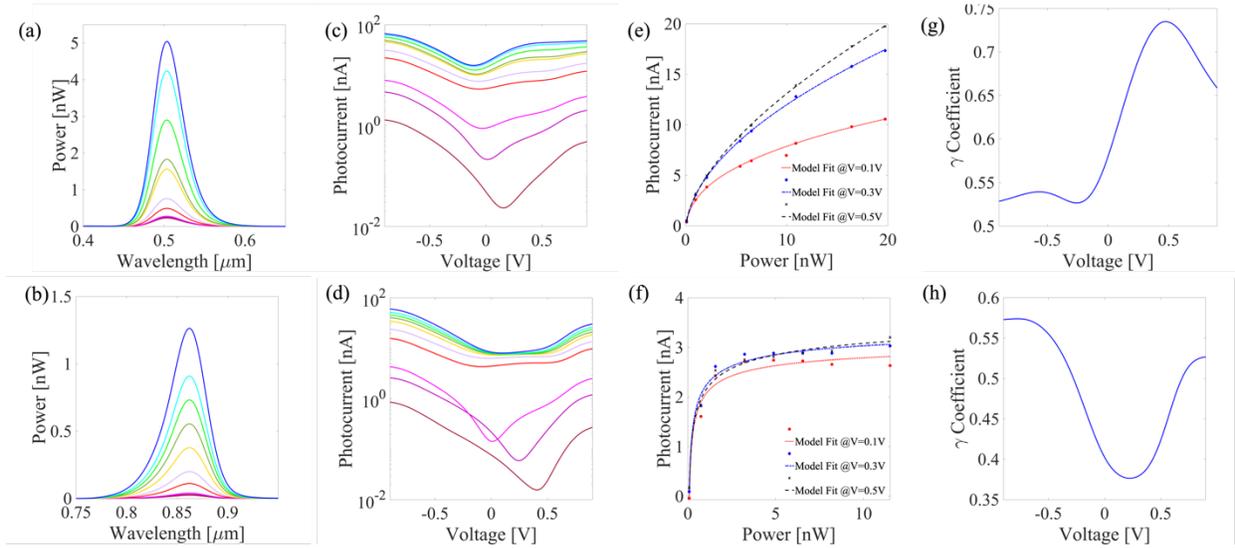

**Figure 3.** Spectral power density of LED sources from the nonlinear training set (a, b), corresponding to the voltage-dependent photocurrent (c, d), with a nonlinear saturation depicted from the measured photocurrents at fixed voltages of 0.1, 0.3 and 0.5 V (black cross, blue diamond and red dot, respectively, fitted to dash, dash-dot and dotted lines) (e, f) that are translated to a voltage-dependent nonlinear coefficient (g, h).

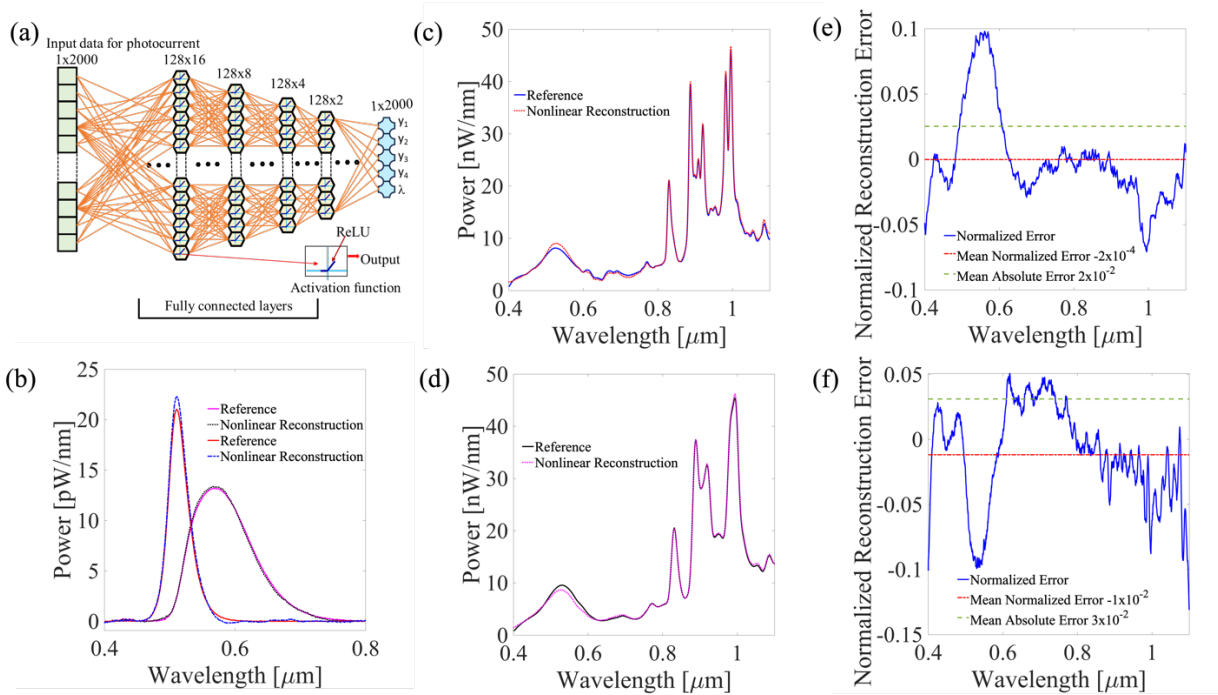

**Figure 4.** Nonlinear reconstruction of power spectra. Schematic of the fully-connected MLP network trained in the encoding process, comprising four hidden layers with a ReLu activation function and input/output vectors of dimension 2,000×1 (a). The reference power spectral density (pink and red lines) and corresponding reconstruction (black dotted and blue dash) of two LED light sources (b). The reference spectrum (blue line) and reconstruction (red dotted) of a color-printed transparency (c), and the same sample reconstructed at a resolution of 1,000x1 input/output vectors (d). (e, f): The reconstruction error corresponding to (c, d) (blue line), the error mean (red dashed) and the absolute error mean (green dot-dashed), respectively.

In the encoding process, pairs of known spectrum and corresponding measured photocurrent vector were utilized for training a multi-layer perceptron (MLP) neural network with four fully connected hidden layers. The hyper-parameter space of the MLP was optimized using 128 batches, consisting of 4 layers with 2048, 1024, 512 and 128 neurons, and a ReLu activation function: $f(x) =$

max $\{0, x\}$ (**Figure 4a**). The dataset was split randomly into a training set (80% of all data) and test set (20%) (see methods section and supplementary text for further details on the training set and procedure). The reconstruction of the power spectrum of two LED sources is demonstrated in **Figure 4b**, showing a relatively low deviation of up to ~5 pW/nm – as compared to a benchtop reference spectrometer (see methods section). This sensitivity results from the training set of **Figure 3a** and **3b**, and **Supplementary Figure 5,** spanning three degrees of order in the power. Furthermore, the resolution of the reconstructed power spectra was evaluated with respect to the dimension of the input/output vectors (photocurrent and spectrum) by reducing the dimension of the power spectrum vectors from $2,000 \times 1$ to $1,000 \times 1$. **Figure 4c** shows the measured and reconstructed spectra of a color-printed polymer transparency (such as the one shown in **Figure 5**), sampled with 2,000 points over the spectral range of 400-1,100 nm. The lower-resolution spectrum of dimension $1,000 \times 1$ is shown in **Figure 4d**. The low-resolution spectrum was decoded with a similar MLP with hidden layers of 1024, 512, 256 and 128 neurons. Interestingly, even upon decreasing the vector size by a factor of two, the power calibration (and dynamic range) of the spectrometer is maintained while the spectral resolution deteriorates, as observed from the broadening and loss of spectral details in **Figure 4d**. Here, the spectral resolution is not defined directly by the reconstruction of the computational spectrometer but based on the resolution of the reference signal that was measured with a tabletop spectrometer. The quality of the computational spectrometer is defined by the reconstruction error relative to the reference signal used to test and match its performance. The normalized reconstruction error, defined as $\mathcal{E} = \frac{\mathcal{P}_{ref} - \mathcal{P}_{rec}}{P} \Delta\lambda$, where $\Delta\lambda$ is the spectral range, $P$ is the total power, $\mathcal{P}_{ref}$ and $\mathcal{P}_{rec}$ are the reference and reconstructed power spectra, respectively. The error is shown in **Figures 4e** and **4f** for the high-and low-resolution cases presented in **Figures 4c** and **4d**, respectively. The mean value of the normalized error and of the absolute error (dash lines) are $-2 \times 10^{-4}$ and $2 \times 10^{2}$, respectively for the high-resolution case of **Figure 4e** and $-10^{-2}$ and $3 \times 10^{-2}$, respectively, for the lower resolution of **Figure 4f**. Clearly, the reconstructed spectra follow the reference, as evident from these small errors.

The ability to image with high spectral resolution within the visible-to-NIR range using a simple portable device has many potential applications in our daily lives. One such common example is the objectivity of colors in vision and imaging, and their dependence upon illumination, known as metamerism. In metamerism, two objects of different color can appear to have the same color or, alternatively, the same object can appear different under varying illumination. The two different filters in Figure 5a appear indistinguishable under fluorescent ambient light. However, with a cell phone flashlight, the difference between these filters becomes evident (**Figure 5b**).

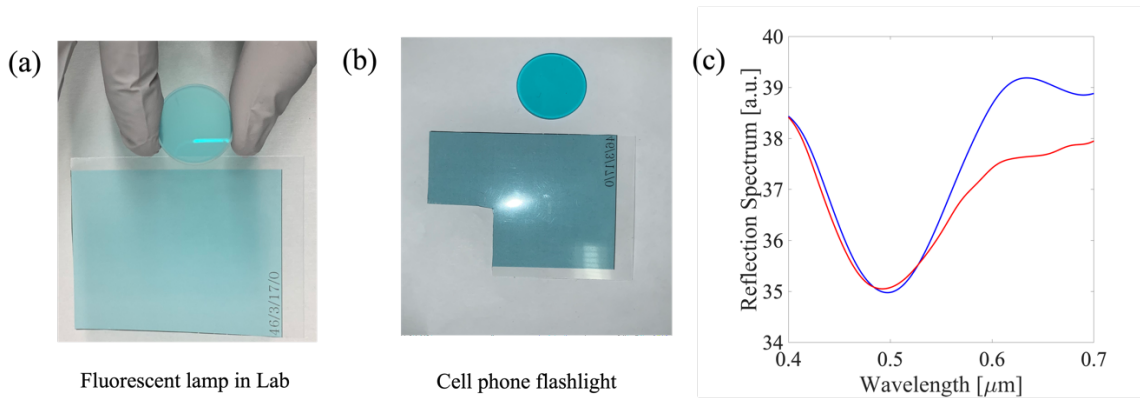

**Figure 5**. A photo of two filters with different color taken and the lighting of a fluorescent lamp (a), the same two filters at the lighting of cell phone flashlight (b) and the reflection spectra corresponding to the two filters as recorded with our nonlinear spectrometer (c).

With our compact nonlinear spectrometer, the reflection spectra of the two filters show a clear difference (**Figure 5c**) and provide a quantitative measure of the true color of the filters, independent of the lighting conditions.

SUMMARY AND CONCLUSIONS

In summary, we have demonstrated a power-calibrated spectrometer based on a single, voltage-tunable GeSe–InSe *p-n* heterojunction device. Using ANNs to decode the nonlinear photocurrent response of this device allows us to achieve high-resolution spectral measurements over the entire visible-to-near-IR range of the optical spectrum. Specifically, with a small device footprint of $\sim 25 \times 25\ \mu m$ and a trained ANN, we were able to reconstruct complex power spectra within the spectral range of 400-1100 nm with accuracy better than 5pW/nm and a spectral resolution of 0.35 nm. Our results pave the way for expanding the role of computational spectroscopy as a viable alternative to traditional optical spectrometry, potentially leading to single-element, on-chip spectrometers for rapid and inexpensive optical sensing.

METHODS

**Device fabrication:** The following steps were taken to fabricate an encapsulated hBN/Graphene/InSe/GeSe vdW heterojunction using the dry transfer method inside a glovebox to prevent contamination and degradation of the exfoliated samples. First, Si/SiO$_2$ substrates were cleaned with deionized water, acetone, and isopropanol in succession, then dried with nitrogen gas. Mechanical exfoliation was used to obtain the desired thickness of hBN, graphene, InSe, and GeSe flakes from their parental crystals on top of Si/SiO$_2$ substrates. Next, the dry-transfer technique was utilized to encapsulate the vertical hetero-junction with hBN, creating a vertical hBN/Gr/InSe/GeSe heterostructure. The successive layers were picked up from a 285 nm Si/SiO$_2$ wafer using a polycarbonate membrane, starting with the top layer of hBN, followed by the graphene flake, InSe, and finally the bottom layer of GeSe. The entire stack was deposited on a clean pre-patterned back electrode (Ti/Au: 5/50 nm) at 150°C and the PC membrane was then dissolved using chloroform.

**Photoresponse Characterization:** All measurements of photocurrent as a function of bias voltage were performed at room temperature ($25 \pm 0.1$ °C) under vacuum conditions at ~$10^{-5}$ Torr. Photocurrent was measured with incident light modulated by a mechanical chopper at frequency of 1 kHz, and with low noise current pre-amplifier (Femto DLPCA-200) and lock-in amplifier (Model SR830). In this photocurrent measurement, the heterojunction was illuminated by seven light-emitting diodes and a Laser Driven Light Source (LDLS) as a white-light source combined with a set of bandpass filters and also with transparency printed filters (see supplementary information for details). The reference spectrum of each light source was measured with a Thermo Fisher Scientific Nicolet-iS50R Fourier Transform Infrared (FTIR) spectrometer connected to an external silicon detector (Thorlabs FDS100) and the spectra were normalized to the silicon detector's calibrated responsivity.

**Computational Spectrum Reconstruction:** The dataset (see Supplementary Information) comprising (i) 10 power spectra acquired from seven LED light sources; (ii) Laser-driven white light source (LDLS) with filter sets: (a) a bandpass filter set (at width of 25 nm) in the spectral range of 400-1100 nm; (b) a bandpass filter set (at width of 40 nm) in the spectral range of 400-1100 nm; (c) a set of homemade filters produced by printing transparencies. All filter sets were spectrally characterized. Overall, 50 filters were used with the white-light source. The data were split randomly into training (80%) and testing (20%) sets. The reconstruction code utilized TensorFlow, sklearn, numpy and scipy software packages.

AUTHOR INFORMATION
* Corresponding author e-mail: Doron.Naveh.biu.ac.il
† The authors contributed equally

COMPETING INTERESTS
The authors declare no competing interest.

ACKNOWLEDGMENTS
D.N., R.D., R.K.Y., E.A., M.P., and A.L. would like to thank the US-Israel Binational Science Foundation (BSF) for supporting this work with grant No. 2021721. D.N., R.D., R.K.Y., E.A., M.P., and A.L would like to thank the European Research Commission for supporting this work under the L2D2 EIC Transition Open No. 101058079. A.R. gratefully acknowledges the National Science Foundation (NSF-BSF 2150562) for support. F.X. and J. L. thanks the National Science Foundation for support through grant 2150561.

AUTHOR CONTRIBUTION
All authors discussed the results and contributed to the manuscript.

# Supplementary Information for

# Nonlinear Self-Calibrated Spectrometer with Single GeSe-InSe Heterojunction Device

Rana Darweesh,[1,2,†] Rajesh Kumar Yadav,[1,2,†] Elior Adler,[1,2] Michal Poplinger,[1,2] Adi Levi,[1,2] Jea-Jung Lee,[3] Amir Leshem,[1] Ashwin Ramasubramaniam,[4,5] Fengnian Xia,[3] and Doron Naveh[1,2,*]

[1] Faculty of Engineering, Bar-Ilan University, Ramat-Gan, Israel 52900.

[2] Institue of Nanotechnology and Advanced Materials, Bar-Ilan University, Ramat-Gan, Israel 52900.

[3] Department of Electrical Engineering, Yale University, New Haven, CT, USA.

[4] Department of Mechanical and Industrial Engineering, University of Massachusetts, Amherst, MA, USA 01003.

[5] Materials Science and Engineering Graduate Program, University of Massachusetts, Amherst, MA 01003, USA.

[†] These authors contributed equally.
[*] Corresponding author: doron.naveh@biu.ac.il

**Contains**

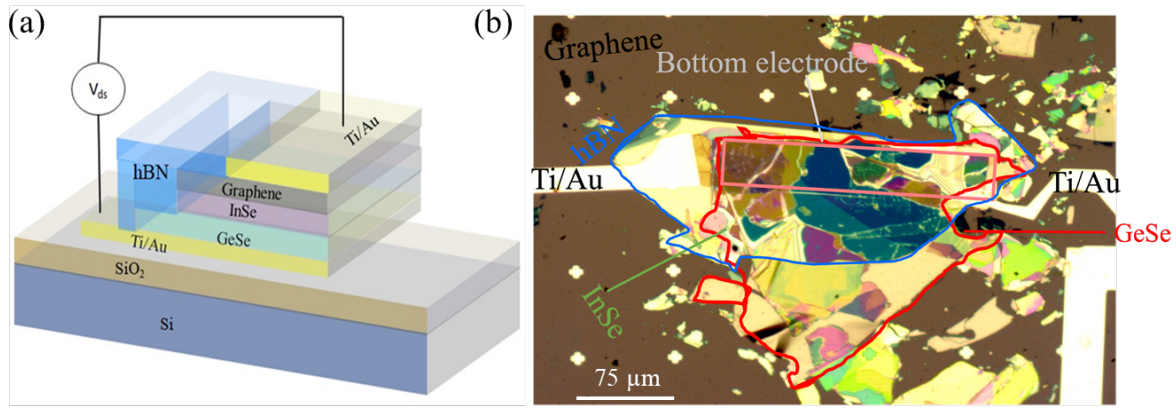

**Supplementary Figure 1. Schematic and optical images of a vertical heterostructure device:** (a) Depiction of the vertical heterostructure composed of hexagonal boron nitride (hBN), graphene, indium selenide (InSe), and germanium selenide (GeSe). This schematic offers a top view representation of the layered arrangement. (b) High-resolution optical image of the vertical heterostructure. The fabrication process adhered to the detailed procedures outlined in the Methods section, ensuring precise layer alignment and material deposition.

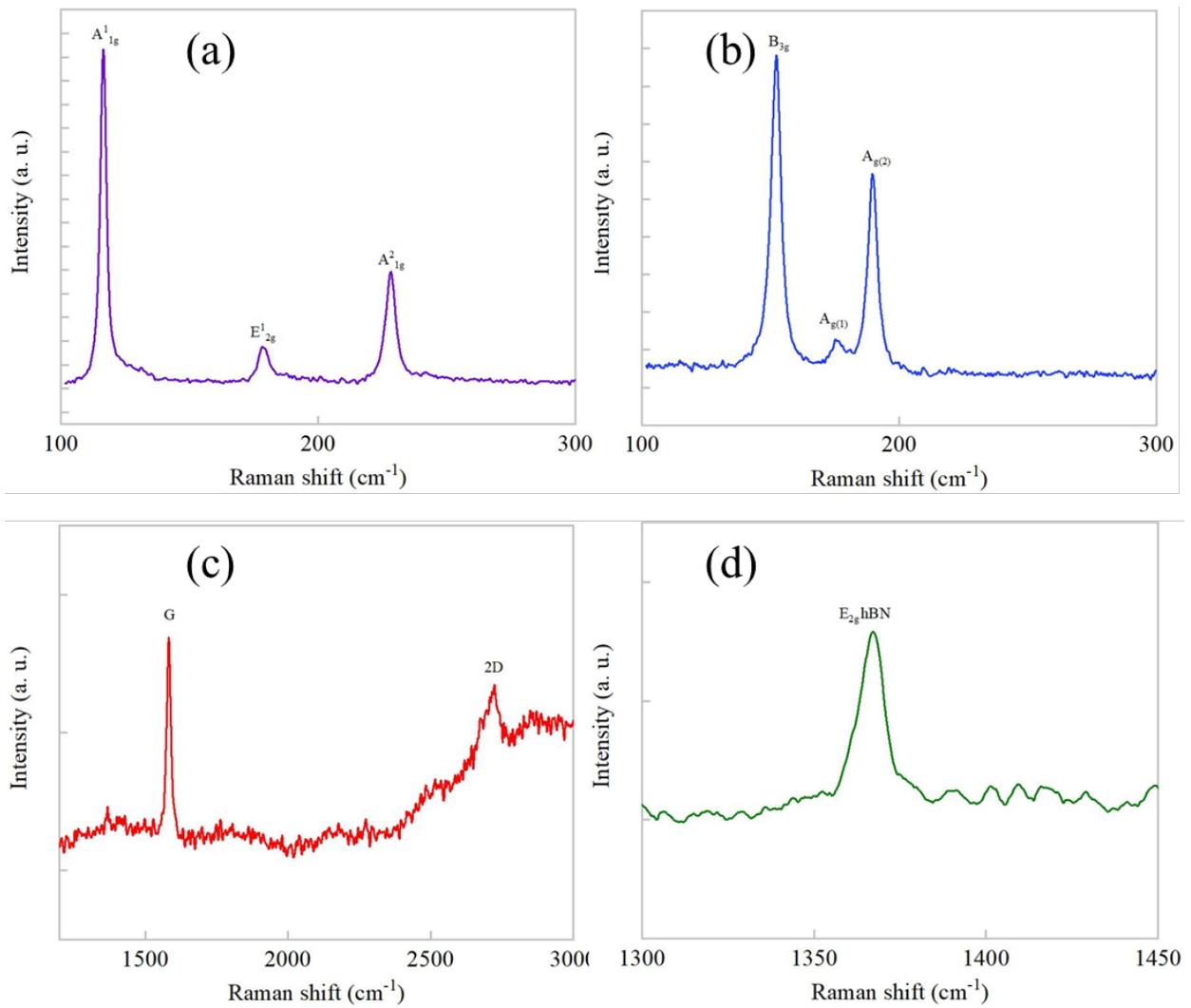

**Supplementary Figure 2. Raman spectra of a heterostructure device after stacking:** Individual (non-overlapping) flakes of (a) InSe, (b) GeSe, (c) Graphene, and (d) hBN.

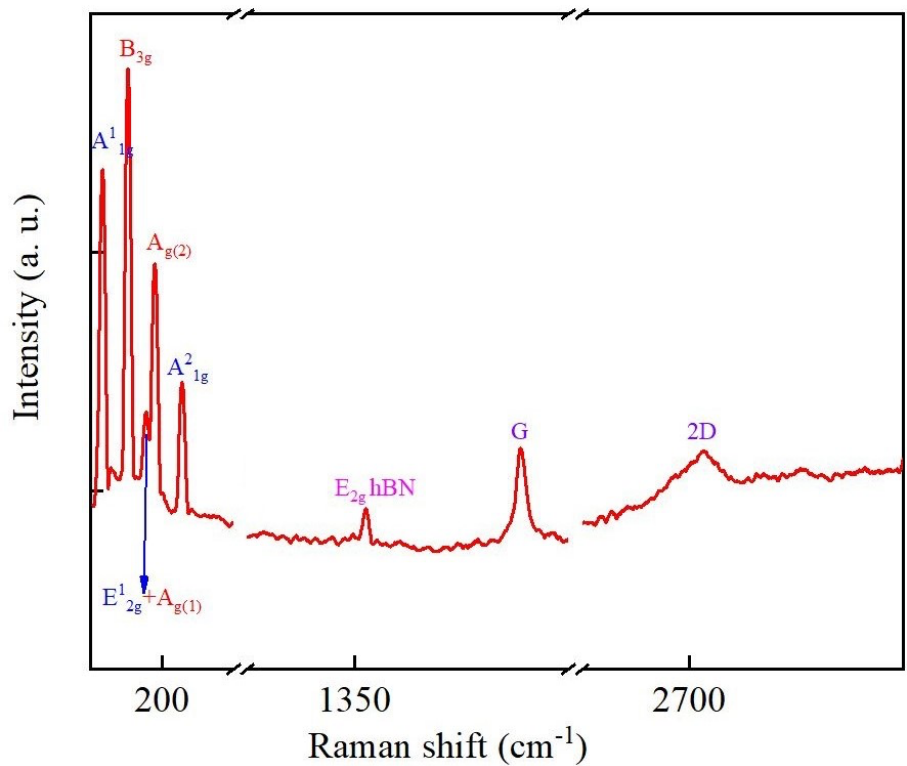

**Supplementary Figure 3.** The Raman spectrum of the heterostructure comprising the spectrometer device exhibits a combination of vibrational modes, including G and 2D modes of graphene, $E_{2g}$ modes of hBN, $A^1_{1g}$, $E^1_{2g}$ and $A^2_{1g}$ modes of InSe and $B_{3g}$, $A_{g\,(1)}$ and $A_{g\,(2)}$ modes of InSe.

**Supplementary Text on Lateral GeSe-InSe Heterostructure Device:**

To verify the operational mechanism of the device in **Figure 1a**, the nature of the Au contact to GeSe and formation of heterojunction at the interface of GeSe-InSe were investigated, as discussed below. A crossbar device (**Supplementary Figure 4a**) was fabricated and showed an Ohmic I-V response (**Supplementary Figure 4b**) for p-type GeSe and the rectified diode behavior for the GeSe-InSe (**Supplementary Figure 4c**).

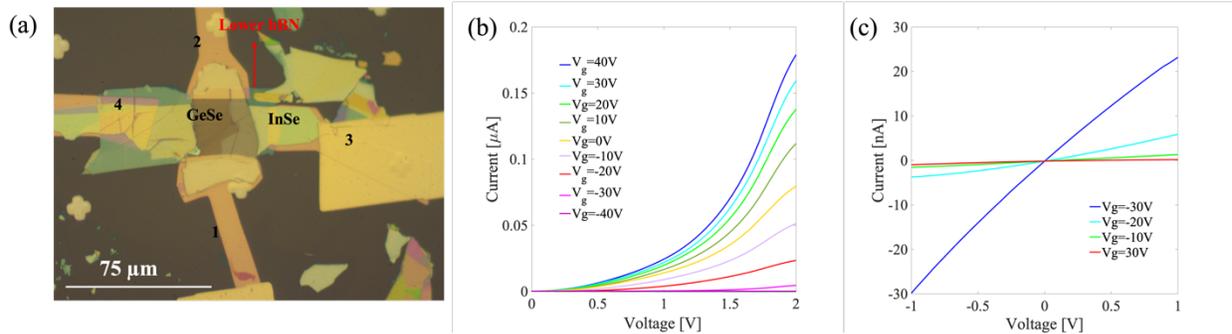

**Supplementary Figure 4. Fabricated lateral device and its current vs. voltage measurements.** (a) High-resolution optical image of a lateral hBN/InSe/GeSe heterostructure. Contacts 1 and 2 were electrically connected to the GeSe flake, while contacts 3 and 4 were linked to the InSe flake. (b) The current-voltage transfer curve under different gate voltages of the lateral device, demonstrating the heterojunction's conductivity arising from the PN junction between GeSe (contact 1) and InSe (contact 3). (c) The current-voltage curve of GeSe (contacts 3 and 4) showing linear behavior.

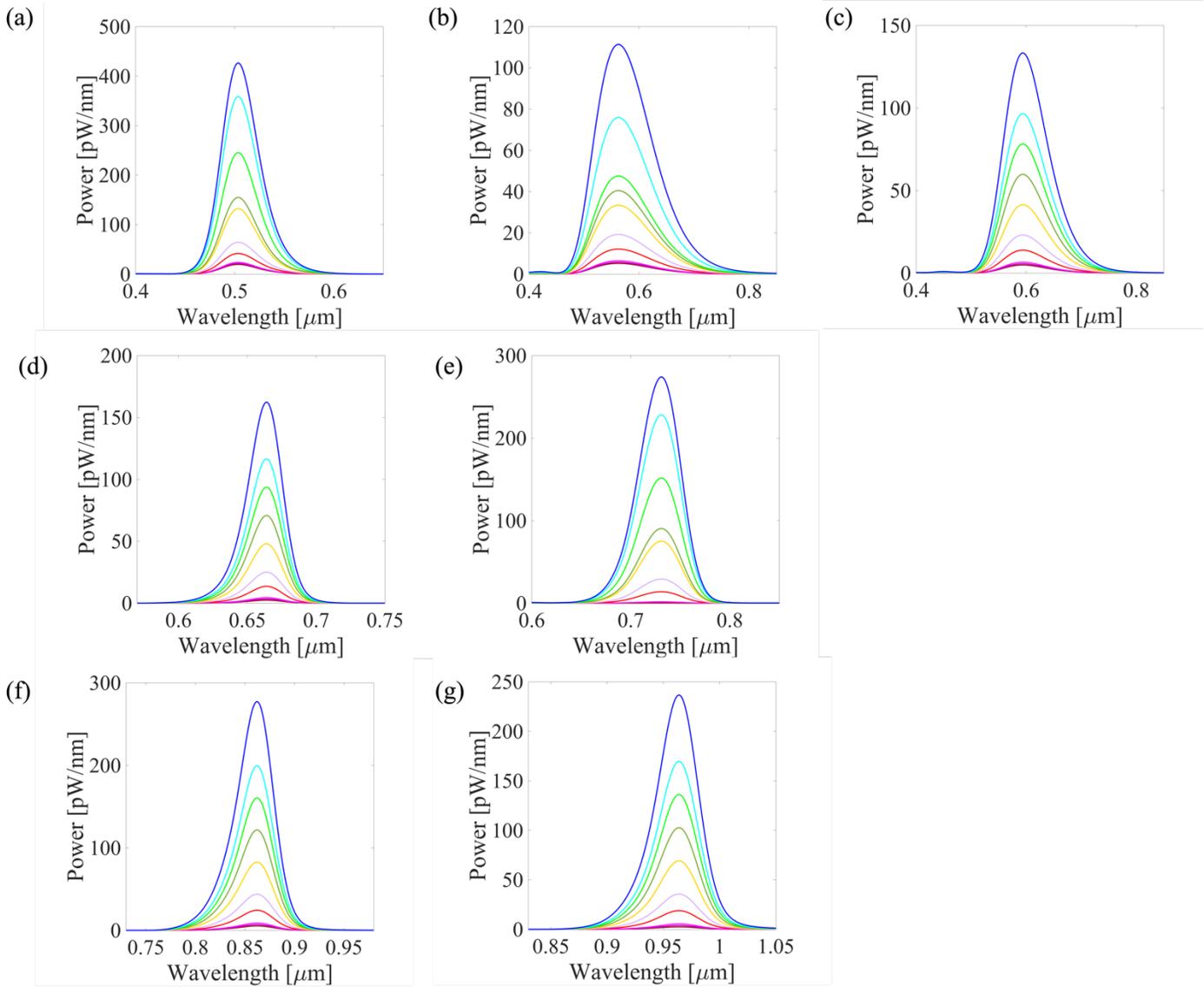

**Supplementary Figure 5. Dataset of the LED spectra used to train the ANN.** The reference spectra of seven LEDs covering the spectral range of the device. The LEDs are centered at wavelengths of (a) 505 nm, (b) 565 nm, (c) 595 nm, (d) 660 nm, (e) 730 nm, (f) 850 nm, and (g) 940 nm. Each LED was measured at ten intensities—highest in blue to lowest in brown.

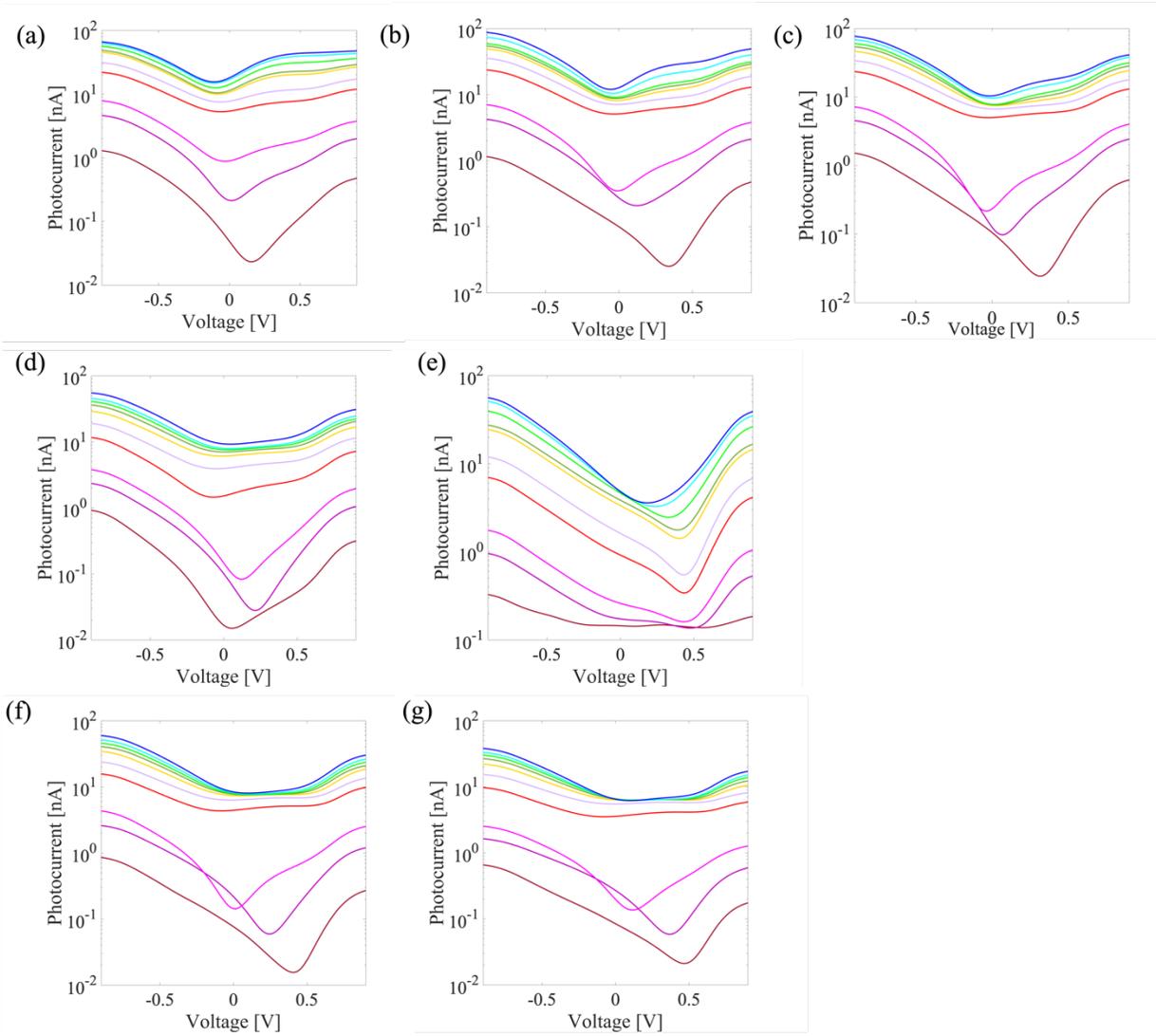

**Supplementary Figure 6. Dataset of the LED photocurrents vs. voltage used to train the ANN.** Photocurrent as function of voltage measured for seven LEDs covering the spectral range of the device. The LEDs are centered at wavelengths of (a) 505 nm, (b) 565 nm, (c) 595 nm, (d) 660 nm, (e) 730 nm, (f) 850 nm, and (g) 940 nm. Each LED was measured at ten intensities—highest in blue to lowest in brown.

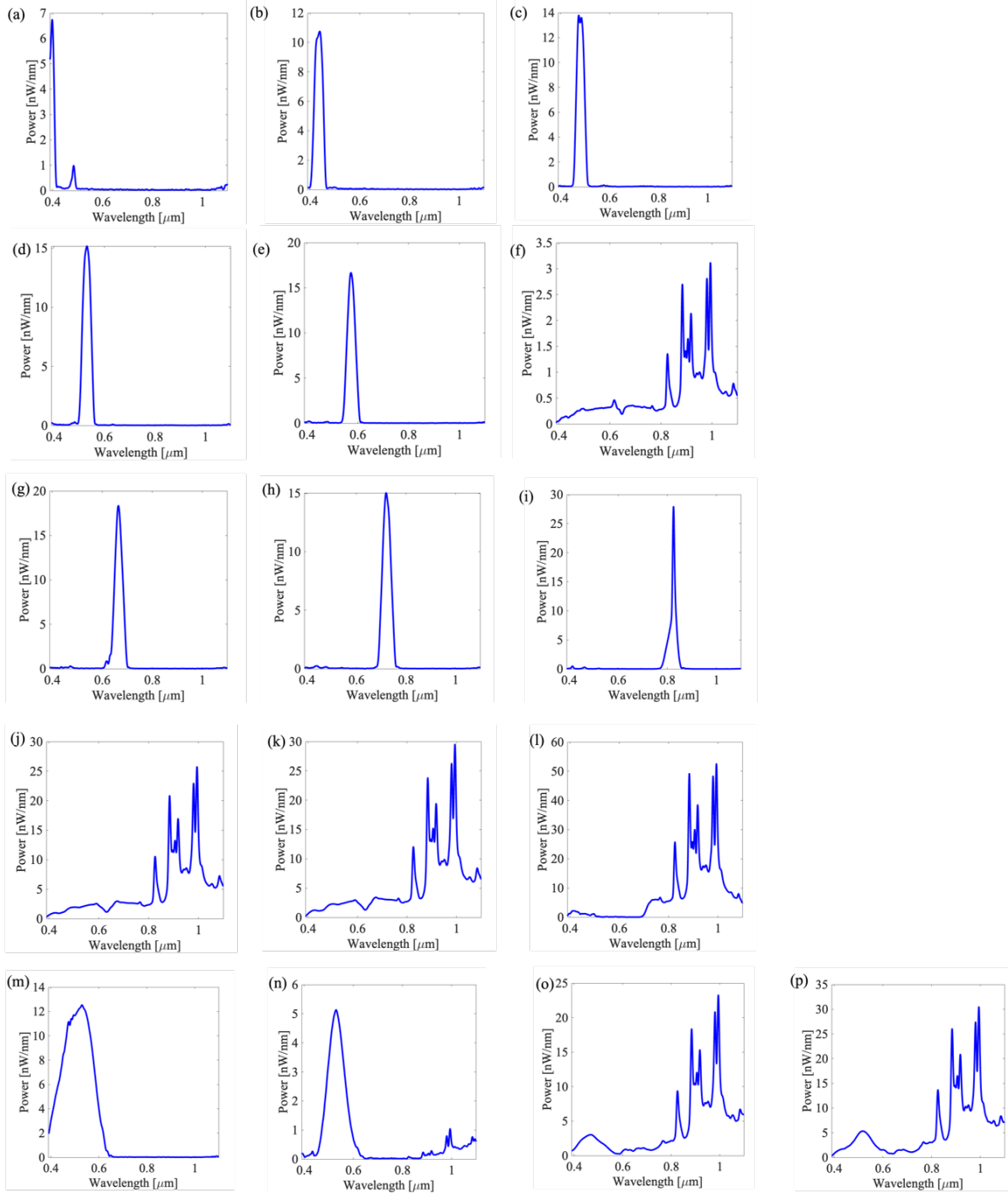

**Supplementary Figure 7. Dataset of the spectra of Laser Driven Light Source (LDLS) with various filters used to train the ANN.** The spectra were measured using Thorlabs bandpass spectral filters, as well as in-house, color-printed polymers.

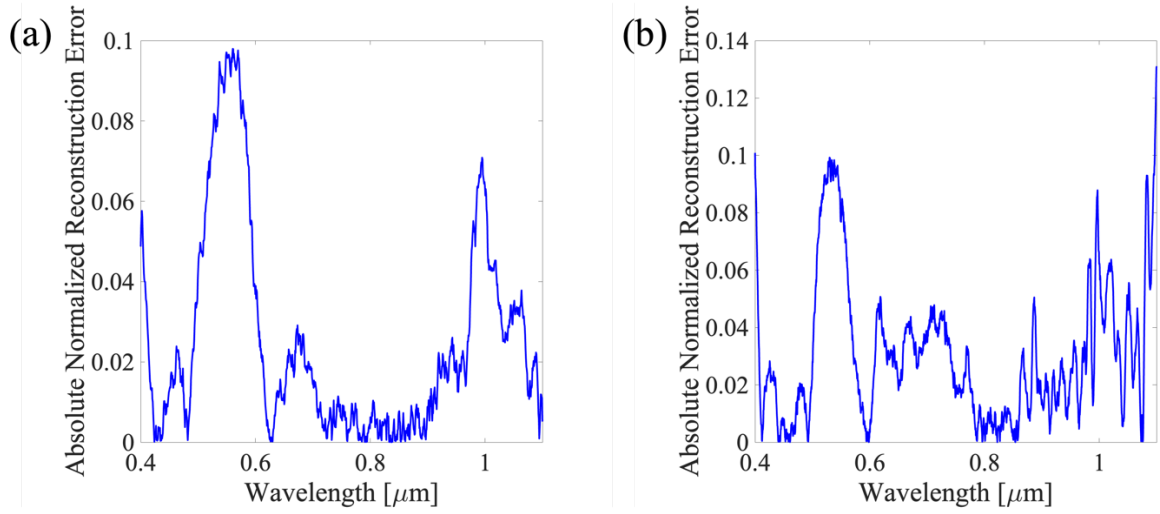

**Supplementary Figure 8. The absolute normalized error of the reconstruction** for the high-resolution (a) and low-resolution (b) cases presented in Figures 4c and 4d, respectively.